# RBF-based meshless boundary knot method and boundary particle method

Wen CHEN[*]

Department of Informatics, University of Oslo, P.O.Box 1080, Blindern, 0316 Oslo, Norway
(E-mail: wenc@ifi.uio.no)

**Classification**: meshfree methods

**Abstract**: This paper is concerned with the two new boundary-type radial basis function collocation schemes, boundary knot method (BKM) and boundary particle method (BPM). The BKM is developed based on the dual reciprocity theorem, while the BKM employs the multiple reciprocity technique. Unlike the method of fundamental solution, the two methods use the non-singular general solution instead of singular fundamental solution to circumvent the controversial artificial boundary outside physical domain. Compared with the boundary element method, both BKM and BPM are meshless, super-convergent, meshfree, integration-free, symmetric, and mathematically simple collocation techniques for general PDE's. In particular, the BPM does not require any inner nodes for inhomogeneous problems. In this study, the accuracy and efficiency of the two methods are numerically demonstrated to some 2D, 3D Helmholtz and convection-diffusion problems under complicated geometries.

## 1.Introduction

In last decade much effort has been devoted to developing a variety of meshless schemes for numerical partial differential equation (PDE) discretization. The driving force behind the scene is that the mesh-based methods such as the standard FEM and BEM often require prohibitively computational effort to mesh or remesh in handling high-dimensional, moving boundary, and complex-shaped boundary problems. Many of the meshfree techniques available now are based on using moving least square strategy (MLS). In most cases, a shadow element is still necessary for numerical integration rather than for function interpolation. Therefore, these methods are not truly meshfree.

Exceptionally, the methods based on radial basis function (RBF) are inherently meshfree due to the fact that the RBF method does not employ the MLS at all and uses the one-dimensional distance variable irrespective of dimensionality of problems. Therefore, the RBF methods are independent of dimensionality and complexity of problem geometry. Nardini and Brebbia[1] in 1982 have actually applied the RBF concept to develop currently popular dual reciprocity BEM (DR-BEM) without a notion of "RBF" and the use of then related advances in multivariate scattered data processing. Only after Kansa's pioneer work [2] in 1990, the research on the RBF method for PDE's has become very active.

Among the existing RBF schemes, the so-called Kansa's method is a domain-type collocation technique, while the method of fundamental solution (MFS) (regular BEM versus singular BEM) is a typical boundary-type RBF collocation methodology. The MFS outperforms the standard BEM in terms of integration free, convergence speed, easy-to-use, and meshfree merits[3]. The main drawback of the MFS is due to use the fictitious boundary outside physical domain. The arbitrariness in the determination of the artificial boundary introduces such troublesome issues as stability and accuracy in dealing with complicated geometry systems and impedes the efficacy of the MFS to practical engineering problems.

Instead of using the singular fundamental solution, Chen and Tanaka[4,5] exploited the non-singular general solution to the approximation of homogeneous solution and removed the controversial artificial boundary in the MFS. The method is called the boundary knot method (BKM). Like the MFS and DRBEM, the BKM also employs the dual reciprocity method to approximate particular solution. For recent development of the BKM see ref. [6]. Some preliminary numerical experiments[4-7] show that the BKM can produce excellent results with relatively a small number of nodes for various linear and nonlinear problems.

---

[*] This research is supported by Norwegian Research Council. The final submission is on 27 Aug. 2001.

On the other hand, in recent years the multiple reciprocity BEM (MR-BEM)[8] has attracted much attention in the BKM community. The method uses high-order fundamental solutions to approximate high-order homogeneous solutions and then get the particular solutions. The advantages and disadvantages of the MRM relative to the DR-BEM are that it does not use inner nodes at all for inhomogeneous problems but that it requires more computing effort. By analogy with the MR-BEM, Chen[6,7] introduced the meshfree boundary particle method (BPM) which combines the RBF and multiple reciprocity principle to formulate a simple and efficient boundary-only meshless collocation method.

The purpose of this study is to introduce the BKM and BPM and testify them to some typical 2D and 3D PDE systems under complicated geometry. The paper is concluded with some remarks.

## 2. Boundary knot method

To clearly illustrate our idea, consider the following example without loss of generality

$$\Re\{u\} = f(x), \quad x \in \Omega, \tag{1}$$

$$u(x) = R(x), \quad x \subset S_u, \tag{2a}$$

$$\frac{\partial u(x)}{\partial n} = N(x), \quad x \subset S_T, \tag{2b}$$

where $\Re$ is differential operator, $x$ means multi-dimensional independent variable, and $n$ is the unit outward normal. The solution of Eq. (1) can be expressed as

$$u = u_h + u_p, \tag{3}$$

where $u_h$ and $u_p$ are the homogeneous and particular solutions, respectively. The latter satisfies

$$\Re\{u_p\} = f(x) \tag{4}$$

but does not necessarily satisfy boundary conditions.

To evaluate the particular solution, the inhomogeneous term is approximated first by

$$f(x) \cong \sum_{j=1}^{N+L} \lambda_j \varphi(r_j), \tag{5}$$

where $\lambda_j$ are the unknown coefficients. $N$ and $L$ are respectively the numbers of knots on the domain and boundary. The use of interior points is usually necessary to guarantee the accuracy and convergence of the BKM solution. $r_j = \|x - x_j\|$ represents the Euclidean distance norm, and $\varphi$ is the radial basis function.

By forcing approximation representation (5) to exactly satisfy Eq. (4) at all nodes, we can uniquely determine

$$\lambda = A_\varphi^{-1}\{f(x_i)\}, \tag{6}$$

where $A_\varphi$ is nonsingular RBF interpolation matrix. Finally, we can get particular solutions at any point by summing localized approximate particular solutions

$$u_p = \sum_{j=1}^{N+L} \lambda_j \phi(\|x - x_j\|), \tag{7}$$

where the RBF $\phi$ is related to the RBF $\varphi$ through operator $\Re$. Substituting Eq. (6) into Eq. (7) yields

$$u_p = \Phi A_\varphi^{-1}\{f(x_i)\}, \tag{8}$$

where $\Phi$ is a known matrix comprised of $\phi(r_{ij})$.

On the other hand, the homogeneous solution $u_h$ has to satisfy both governing equation and boundary conditions. Unlike the dual reciprocity BEM[1] and MFS[3] using the singular fundamental solution, the BKM[4-8] approximates $u_h$ by means of nonsingular general solution, namely,

$$u_h(x) = \sum_{k=1}^{L} \alpha_k u^{\#}(r_k), \qquad (9)$$

where $k$ is the index of source points on boundary; $u^{\#}$ is the nonsingular general solution of operator $\Re$. $\alpha_k$ are the desired coefficients. Collocating Eqs. (1) and (2a,b) at all boundary and interior knots in terms of representation (8) and (9), we have the unsymmetric BKM schemes. For the sake of brevity, the respective details are omitted here[4-7]. In order to get symmetric BKM scheme for self-adjoint operators, we modify the BKM approximate expression (9) to homogeneous solution $u_h$ as

$$u_h(x) = \sum_{s=1}^{L_d} a_s u^{\#}(r_s) - \sum_{s=L_d+1}^{L_d+L_N} a_s \frac{\partial u^{\#}(r_s)}{\partial n}, \qquad (10)$$

where $n$ is the unit outward normal as in boundary condition (2b), and $L_d$ and $L_N$ are respectively the numbers of knots at the Dirichlet and Neumann boundary surfaces. The minus sign associated with the second term is due to the fact that the Neumann condition of the first order derivative is not self-adjoint. In terms of expression (10), the collocation analogue equations (1a) and (2a,b) are written as

$$\sum_{s=1}^{L_d} a_s u^{\#}(r_{is}) - \sum_{s=L_d+1}^{L_d+L_N} a_s \frac{\partial u^{\#}(r_{is})}{\partial n} = R(x_i) - u_p(x_i), \qquad (11)$$

$$\sum_{s=1}^{L_d} a_s \frac{\partial u^{\#}(r_{js})}{\partial n} - \sum_{s=L_d+1}^{L_d+L_N} a_s \frac{\partial^2 u^{\#}(r_{js})}{\partial n^2} = N(x_j) - \frac{\partial u_p(x_j)}{\partial n}, \qquad (12)$$

$$\sum_{s=1}^{L_d} a_s u^{\#}(r_{ls}) - \sum_{s=L_d+1}^{L_d+L_N} a_s \frac{\partial u^{\#}(r_{ls})}{\partial n} = u_l - u_p(x_l). \qquad (13)$$

The system matrix of the above equations is symmetric if operator $\Re\{\}$ is self-adjoint. Note that $i$, $s$ and $j$ are reciprocal indices of Dirichlet ($S_u$) and Neumann boundary ($S_\Gamma$) nodes. $l$ indicates response knots inside domain $\Omega$. After the solution of the above simultaneous algebraic equations, we can employ the obtained expansion coefficients $\alpha$ and inner knot solutions $u_l$ to calculate the BKM solution at any knot. It is stressed here that the MFS could not produce the symmetric interpolation matrix in any way.

The present form of the BKM uses the expansion coefficient rather than the direct physical variable in the approximation of boundary value. Therefore, such BKM is called as the indirect BKM. Chen et al. [6,7] also gave the direct BKM with physical variable as basic variable.

## 3. Boundary particle methods

Just like the DR-BEM, the interior nodes are usually necessary in the BKM for inhomogeneous problems. A rival to the DR-BEM is the MR-BEM[8], which applies the multiple reciprocity principle to circumvent the domain integral without using any inner node. In this section, we will develop a boundary-only RBF scheme called the boundary particle method [6,7] with the multiple reciprocity principle.

The multiple reciprocity method assumes that the particular solution in Eq. (3) can be approximated by higher-order homogeneous solution, namely,

$$u = u_h^0 + u_p^0 = u_h^0 + \sum_{m=1}^{\infty} u_h^m, \qquad (14)$$

where superscript $m$ is the order index of homogeneous solution. $u_h^0$ and $u_p^0$ are equivalent to homogeneous solution $u_h$ and particular solution $u_p$ in Eq. (3). Through an incremental differentiation operation via operator $\Re\{\}$, we have successively higher order differential equations:

$$\begin{cases} u_h^0(x_i) = R(x_i) - u_p^0(x_i) \\ \dfrac{\partial u_h^0(x_j)}{\partial n} = N(x_j) - \dfrac{\partial u_p^0(x_j)}{\partial n} \end{cases}, \qquad (15a)$$

$$\begin{cases} \Re^0\{u_h^1(x_i)\} = f(x_i) - \Re^0\{u_p^1(x_i)\} \\ \dfrac{\partial \Re^0\{u_h^1(x_j)\}}{\partial n} = \dfrac{\partial (f(x_j) - \Re^0\{u_p^1(x_j)\})}{\partial n} \end{cases}, \qquad (15b)$$

$$\begin{cases} \Re^{n-1}\{u_h^n(x_i)\} = \Re^{n-2}\{f(x_i)\} - \Re^{n-1}\{u_p^n(x_i)\} \\ \dfrac{\partial \Re^{n-1}\{u_h^n(x_j)\}}{\partial n} = \dfrac{\partial(\Re^{n-2}\{f(x_j)\} - \Re^{n-1}\{u_p^n(x_j)\})}{\partial n} \end{cases}, \quad n=2,3,\ldots, \tag{15c}$$

where $\Re^n\{\}$ denotes the *n*-th order operator $\Re\{\}$, say $\Re^1\{\}=\Re\Re^0\{\}$, and $\Re^0\{\}=\Re\{\}$, $i$ and j are respectively Dirichlet and Neumann boundary knots. $u_p^n$ is the *n*-th order of particular solution defined as

$$u_p^n = \sum_{m=n+1}^{\infty} u_h^m. \tag{16}$$

The m-order homogeneous solution is approximated by

$$u_h^m(x) = \sum_{k=1}^{L} \beta_k^m u_m^\#(r_k), \tag{17}$$

where $L$ is the number of boundary nodes, and $u_m^\#$ is the corresponding *m*-th order fundamental or general solutions. Collocating boundary equations (15a,b,c) only on boundary nodes, we have the boundary discretization equations

$$\left.\begin{array}{l} \sum_{k=1}^{L} \beta_k^0 u^0(r_{ik}) = R(x_i) - u_p^0(x_i) \\ \sum_{k=1}^{L} \beta_k^0 \dfrac{\partial u^0(r_{jk})}{\partial n} = N(x_j) - \dfrac{\partial u_p^0(x_j)}{\partial n} \end{array}\right\} = b^0, \tag{18a}$$

$$\left.\begin{array}{l} \sum_{k=1}^{L} \beta_k^1 \Re^0\{u_1^\#(r_{ik})\} = f(x_i) - \Re^0\{u_p^1(x_i)\} \\ \sum_{k=1}^{L} \beta_k^1 \dfrac{\partial \Re^0\{u_1^\#(r_{jk})\}}{\partial n} = \dfrac{\partial(f(x_j) - \Re^0\{u_p^1(x_j)\})}{\partial n} \end{array}\right\} = b^1, \tag{18b}$$

$$\left.\begin{array}{l} \sum_{k=1}^{L} \beta_k^n \Re^{n-1}\{u_n^\#(r_{ik})\} = \Re^{n-2}\{f(x_i)\} - \Re^{n-1}\{u_p^n(x_i)\} \\ \sum_{k=1}^{L} \beta_k^n \dfrac{\partial \Re^{n-1}\{u_n^\#(r_{jk})\}}{\partial n} = \dfrac{\partial(\Re^{n-2}\{f(x_j)\} - \Re^{n-1}\{u_p^n(x_j)\})}{\partial n} \end{array}\right\} = b^n, \quad n=2,3,\ldots. \tag{18c}$$

In terms of the MRM, the successive process is truncated at some order *M*, namely, let

$$\Re^{M-1}\{u_p^M\} = 0. \tag{19}$$

The practical solution procedure is a reversal recursive process:

$$\beta_k^M \to \beta_k^{M-1} \to \cdots \to \beta_k^0. \tag{20}$$

It is noted that due to

$$\Re^{n-1}\{u_h^n(r_k)\} = u_h^0(r_k), \tag{21}$$

the coefficient matrices of all successive equation are the same, i.e.

$$Q\beta_k^n = b^n, \quad n=M,M-1,\ldots,1,0, \tag{22}$$

Thus, the LU decomposition algorithm is suitable for this task. Finally, the solution is given by

$$u(x_i) = \sum_{n=0}^{M} \sum_{k=1}^{L} \beta_k^n u_n^\#(r_{ik}). \tag{23}$$

The BPM can use either singular fundamental solution or nonsingular general solution, respectively relative to the MFS and BKM. It is noted that the BPM with *M*=1 degenerates into the BKM or MFS without using the inner nodes. The only difference between the BKM (MFS) and BPM lies in how to evaluate the particular solution. The former applies the dual reciprocity principle, while the latter employs the multiple reciprocity principle. The advantage of the BPM over the BKM is that it dose not require interior nodes which may be especially attractive in such problems as moving boundary, inverse problems, and exterior problems. However, the BPM may be more mathematically complicated and computationally costly due to the iterative use of higher-order fundamental or general solutions. It is expected that like the MRM[8], the truncated order *M* in the BPM may not be large (usually two or three orders) in a variety of practical uses. The above form of the BPM is unsymmetric. It is rather straightforward to derive the symmetric BPM scheme[6,7] by replacing Eq. (17) with a representation similar to Eq. (10) and thus, matrix $Q$ in Eq. (22) will be symmetric if operator $\Re\{\}$ is self-adjoint.

## 3. Numerical experiments

The 2D and 3D irregular geometries tested are illustrated in Figs. 1 and 2, which respectively include a triangle cut-out and a two-ball (radii=1, center distance=$\sqrt{2}$) cavity. Unless specified Neumann boundary conditions shown in Fig. 1 and on $x=0$ surface in Fig. 2, the otherwise boundary are all Dirichlet type. The BKM employs 9 inner nodes for 2D inhomogeneous case as shown by small crosses in Fig. 1. Equally spaced knots were applied on the boundary except on two-ball surface where the random knots were employed. The $L_2$ norms of relative errors are calculated by the numerical solutions at 364 nodes for 2D and 500 nodes for 3D. The absolute error is taken as the relative error if the absolute value of the solution is less than 0.001. Note that different nodes are used for BKM and BPM coefficients and for $L_2$ norm of relative errors.

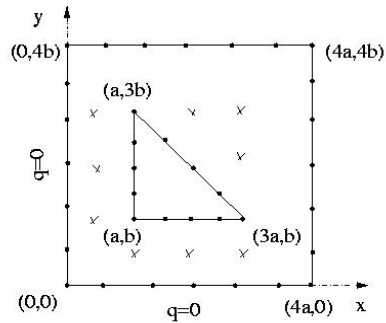
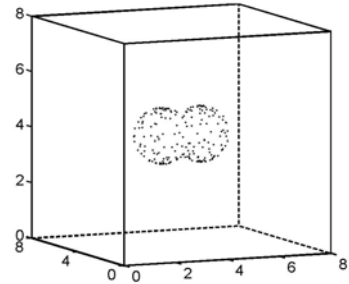

Fig. 1. Configuration of a square with a trigeometric cutout       Fig. 2. A cube with a two-ball cavity

Equations of Helmholtz and diffusion-reaction
$$\nabla^2 u + \gamma^2 u = f(x), \qquad \nabla^2 u - \tau^2 u = q(x) \qquad (24a,b)$$
and convection-diffusion
$$D\nabla^2 u(x) - v \bullet \nabla u(x) - \kappa u(x) = g(x), \qquad (25)$$
are examined, where $v$ denotes a velocity vector, $\tau$ is the Thiele parameter, $D$ is the diffusivity coefficient, $\kappa$ represents the reaction coefficient. The accurate solutions are
$$u = x^2 \sin x \cos y \qquad (26)$$
for 2D inhomogeneous Helmholtz problem ($\gamma = \sqrt{2}$) and
$$u = \sin x \cos y \cos z \qquad \text{and} \qquad u = e^{-\sigma x} + e^{-\sigma y} + e^{-\sigma z} \qquad (27a,b)$$
for 3D homogeneous Helmholtz ($\gamma = \sqrt{3}$) and convection-diffusion ($v_x=v_y=v_z=-\sigma$, $\kappa=0$, Peclect number is 24 for $\sigma=1$ and 480 for $\sigma=20$) problems. The corresponding inhomogeneous function $f(x)$, $g(x)$ and boundary conditions can be derived accordingly. The high-order general solutions of Helmholtz, diffusion, and convection-diffusion problems are respectively given by
$$u_m^\#(r) = A_m(\gamma r)^{-n/2+1+m} J_{n/2-1+m}(\gamma r), \qquad u_m^\#(r) = A_m(\tau r)^{-n/2+1+m} I_{n/2-1+m}(\tau r), \qquad (28)$$
$$u_m^\#(r) = A_m(\mu r)^{-n/2+1+m} e^{\frac{v \cdot r}{2D}} I_{n/2-1+m}(\mu r), \qquad n \geq 2, \qquad (29)$$
where $n$ is the dimension of the problem; $m$ denotes the order of general solution; $J$ and $I$ respectively represent the Bessel and modified Bessel function of the first kind; $A_m = A_{m-1}/(2*m*\gamma^2)$, $A_0=1$.
$$\mu = \left[ \left( |v|/2D \right)^2 + \kappa/D \right]^{\frac{1}{2}}. \qquad (30)$$

In this study, the second-order high order general solutions are used as the radial basis function in the BKM approximation of particular solution. The experimental results are displayed in Tables 1 and 2. It is found that both methods produce very accurate solutions with a small number of nodes for inhomogeneous Helmholtz problems. Like the DR-BEM, the BKM is somewhat sensitive to inner node locations and number. In the 3D case relative to 2D case, more nodes are used due to not only one more dimension but also larger domain and more complicated boundary shape. It is expected that compared with other numerical techniques, the BKM and BPM become more efficient for higher-dimensional complex-shape geometry problems since the general solutions of high-dimensional operators are simpler and radial basis function is independent of dimensionality and geometry complexity.

Table 1. $L_2$ norm of relative errors for 2D inhomogeneous Helmholtz problems by the BKM and BPM (numbers inside parentheses indicate boundary plus inner nodes used by the respective method ).

| BKM (26+9) | BKM (33+9) | BPM (26) | BPM (33) |
|---|---|---|---|
| 1.9e-3 | 9.3e-5 | 2.7e-3 | 6.8e-4 |

Table 2. $L_2$ norm of relative errors for 3D homogeneous Helmholtz and convection-diffusion problems by the BKM.

| Helmholtz | | Convection-diffusion ($\sigma=1$) | | Convection-diffusion ($\sigma=20$) | |
|---|---|---|---|---|---|
| 4.6e-3 (298) | 1.7e-4 (466) | 9.0e-3 (136) | 2.2e-3 (298) | 8.8e-15 (136) | 6.8e-15 (178) |

## 4. Remarks

The BKM and BPM circumvent the troublesome singular integral inherent in the BEM and are very easy to learn and program. It is noted that unlike the MR-BEM, the BPM does not need to generate more than one interpolation matrix, which tremendously reduces the computing effort and storage. In addition, both schemes are essentially meshfree, spectral convergence and symmetric technique. Similar to the comparisons between the DR-BEM and MR-BEM, the BKM may be mathematically simpler and more generally applicable than the BPM, while the latter has advantage not requiring inner nodes for inhomogeneous problems and is very suitable for problems whose higher-order homogeneous solution quickly tends to zero. For a complete description and references of the BKM and BPM see refs. [4-7].

## References

1. Nardini, D. and Brebbia, C.A., A new approach to free vibration analysis using boundary elements, *Applied Mathematical Modeling*, 1983, 7: 157-162.
2. Kansa, E.J. Multiquadrics: A scattered data approximation scheme with applications to computational fluid-dynamics. *Comput. Math. Appl*. 1990, 19: 147-161.
3. Golberg, M.A. and Chen, C.S., The method of fundamental solutions for potential, Helmholtz and diffusion problems. In *Boundary Integral Methods - Numerical and Mathematical Aspects,* (Ed. by M.A. Golberg), pp. 103-176, Comput. Mech. Publ., 1998.
4. Chen, W. and Tanaka, M. A meshless, exponential convergence, integration-free, and boundary-only RBF technique. *Comput. Math. Appl.,* (in press), 2001.
5. Chen, W. and Tanaka, M. New Insights into Boundary-only and Domain-type RBF Methods. *Int. J. Nonlinear Sci. & Numer. Simulation,* 2000, 1(3): 145-151.
6. Chen, W. Several new domain-type and boundary-type numerical discretization schemes with radial basis function. *ACM Computing Research Repository*, http://xxx.lanl.gov/abs/cs.NA/0104018, April, 2001.
7. Chen, W. New RBF collocation schemes and their applications. *Int. Workshop Meshfree Methods for Partial Differential Equations,* Bonn, Germany, Sept. 2001.
8. A.J. Nowak and A.C. Neves (ed.). *The Multiple Reciprocity Boundary Element Method*. Comput. Mech. Publ., Southampton, UK, 1994.